\documentclass[runningheads]{llncs}
\usepackage{graphicx}
\usepackage{cite}
\usepackage{amsmath,amssymb,amsfonts}
\usepackage{algorithmic}
\usepackage{textcomp}
\usepackage{xcolor}
\def\BibTeX{{\rm B\kern-.05em{\sc i\kern-.025em b}\kern-.08em
    T\kern-.1667em\lower.7ex\hbox{E}\kern-.125emX}}

\usepackage[textsize=tiny]{todonotes}
\usepackage{graphicx}
\graphicspath{ {./images/} }
\usepackage{csquotes}
\usepackage{threeparttable}
\usepackage{pifont}
\usepackage{tikz}  
\usetikzlibrary{positioning, shadows, shapes.geometric, shapes.symbols, calc, fit}
\usepackage{wrapfig}   
\usepackage{pgfplots}
\usepackage{pgfplotstable}
\pgfplotsset{compat=1.17}

\usepackage{graphicx}
\usepackage{caption}
\usepackage{booktabs}
\usepackage{comment}
\usepackage[]{xcolor}
\usepackage{multicol}
\usepackage{floatrow}
\usepackage{textpos}
\usepackage{framed}
\newfloatcommand{capbtabbox}{table}[][\FBwidth]

\usepackage[]{hyperref}

\begin{document}
\author{Elias Grünewald \orcidID{0000-0001-9076-9240} \and Leonard Schurbert}
\authorrunning{F. Author et al.}
\institute{Technische Universität Berlin, Information Systems Engineering, Germany 
\email{\{gruenewald, schurbert\}@tu-berlin.de}}

\authorrunning{Anonymous author(s)}
\authorrunning{E. Grünewald and L. Schurbert}

\title{Scalable Discovery and Continuous Inventory of Personal Data at Rest in Cloud Native Systems}
\titlerunning{Scalable Discovery and Continuous Inventory of Personal Data at Rest}

\maketitle  %

\begin{abstract}
    Cloud native systems are processing large amounts of personal data through numerous and possibly multi-paradigmatic data stores (e.g., relational and non-relational databases). From a privacy engineering perspective, a core challenge is to keep track of all exact locations, where personal data is being stored, as required by regulatory frameworks such as the European General Data Protection Regulation. In this paper, we present \textsc{Teiresias}, comprising i) a workflow pattern for scalable discovery of personal data at rest, and ii) a cloud native system architecture and open source prototype implementation of said workflow pattern. To this end, we enable a continuous inventory of personal data featuring transparency and accountability following DevOps/DevPrivOps practices. In particular, we scope version-controlled Infrastructure as Code definitions, cloud-based storages, and how to integrate the process into CI/CD pipelines. Thereafter, we provide iii) a comparative performance evaluation demonstrating both appropriate execution times for real-world settings, and a promising personal data detection accuracy outperforming existing proprietary tools in public clouds.


\end{abstract}

\begin{textblock*}{1.7\textwidth}(-4.3cm,-16.5cm) %
\begin{center}
\begin{framed}
    \textit{Preprint (2022-09-09) before final copy-editing of an accepted peer-reviewed paper to appear in the\\ Proceedings of the \textbf{20\textsuperscript{th} International Conference on Service-Oriented Computing (ICSOC'22)}.}\\ The Version of Record can be found here: \url{https://link.springer.com/conference/icsoc}
\end{framed}
\end{center}
\end{textblock*}

\section{Introduction}

\noindent The European General Data Protection Regulation (GDPR) %
or, similarly, the California Consumer Privacy Act (CCPA)
define strong regulatory frameworks following the principle Privacy\footnote{For the sake of simplicity, we use the terms \textit{privacy} and \textit{data protection} interchangeably, being aware of their different notions in other contexts.}
by Design and by Default (PbD). At the same time, various services collect personal data from countless data subjects and enterprises face the challenges of aligning to all regulatory obligations to avoid severe fines. In particular, data controllers are required to establish technical and organizational measures as safeguards against potential misuse or data breaches. Supervisory authorities are also expanding their activities to audit data controllers and processors \cite{ruohonen2021gdpr}. Meanwhile, cloud native systems follow polyglot microservice architectures and multi-cloud strategies and are therefore especially hard to account for personal data. Due to their inherent complexity, they lack transparency and because of their evolutionary development, they might contradict present accountability requirements.

In practice, personal data are found in multi-paradigmatic storage and processing settings (e.g., SQL, NoSQL). Effective means need to be found that identify common patterns or context-specific indicators of personal data as such in the vast amount of present data at rest. For instance, the GDPR requires comprehensive records of processing activities (RoPA, inventory) according to Art. 30.%
Such a procedure consists of technical and organizational measures, which, in turn, have to take into account the state of the art (Art. 25).
So far, we observe laborious and primarily manual tasks of information collection and documentation, often inflexibly supported by simplistic spreadsheets or burdensome written documents \cite{ropa}. Consequently, we identify a collective need for what data protection officers and supervisory authorities need to enhance transparency and accountability of large-scale cloud native systems: A scalable discovery and continuously updated inventory of personal data. Furthermore, we observe the need for tools to guarantee data protection at runtime to meet the prevailing software development and operations (DevOps) lifecycle of constantly evolving systems and not only ex-ante assumptions. From that follows, the technical scope should align with both the regulatory obligations and the development practice. A key research question here is: \textit{How to (i) discover personal data in large-scale cloud native systems and (ii) how to inventory respective findings?}  

We herein present the (to the best of our knowledge) first model, architecture, and implementation that jointly leverages Infrastructure as Code definitions, multi-paradigmatic data stores, and CI/CD pipelines in cloud native systems to inventory personal data at rest. To this end, we provide in this paper:

\begin{itemize}
    \item A workflow pattern for the scalable discovery of personal data in cloud native systems.
    \item \textsc{Teiresisas}, an architecture and open source prototype implementation of said workflow pattern.
    \item An experimental evaluation of our approach in comparison to two widely used baseline systems.
\end{itemize}

Therefore, the remainder of this paper is structured as follows:
In sec.~\ref{sec:background} we provide relevant background and related work.
Thereafter, in sec.~\ref{sec:general-approach} we present the general approach for a scalable and continuous inventory procedure.
We elaborate on the implementation in sec.~\ref{sec:implementation}.
On this basis, our approach is evaluated in sec.~\ref{sec:evaluation}.
In sec.~\ref{sec:limitations} we discuss the current limitations.
Finally, sec.~\ref{sec:conclusion} concludes.

\section{Background and Related work} \label{sec:background}

\subsection{Personal Data in Cloud Native Systems}

Hereinafter, we refer to personal data as defined in Art. 4(1) which states \enquote{‘personal data’ means any information relating to an identified or identifiable natural person (‘data subject’)}. Looking at real-world data sets, this classification is, however, often far from trivial \cite{finck2020they}. %
The presence of personal data can be constituted by existing field such as a data subject's name, (email) address, social security number, (under certain circumstances) IP address \cite{zuiderveen2017breyer} or more complex data structures such as social media profiles, location data, personal preferences, health records, and many others. Usually, these data are stored in relational or non-relational (NoSQL) databases, e.g., as basic \emph{string} values or more complex objects.

From a legal perspective, several obligations need to be implemented by the data controller as imposed by the GDPR. For this paper, four guiding privacy principles are central:
First, For the processing of personal data the \textit{transparency} principle according to Art. 5(1a) applies. This implies the controller needs to provide detailed transparency information to signal categories of personal data being collected, their retention time, legal basis, purposes, third country transfers, and many more (%
see, e.g., \cite{grunewald2021tilt}). 
Second, Art. 5(1f) requires appropriate \textit{security} measures (incl. integrity and confidentiality) against unauthorized or unlawful processing or accidental loss (cf.~sec.~\ref{ssec:dlp}).  
Third, the controller needs to be able to demonstrate compliance with (not only) the aforementioned principles, i.e. the responsibility and liability (see also Recital 74) for \textit{accountability} (Art. 5(2)). To this end, data controllers shall maintain a record of processing activities (Art. 30) with a special focus on the above-mentioned security of processing (Art. 32). While carrying out Data Protection Impact Assessments (DPIAs), the risks associated with the processing of personal data shall be determined and regularly reviewed (Art. 35 GDPR).
Fourth, the overarching principle of privacy by design and by default (Art. 25) needs to be taken into account. It applies to all obligations laid out in the GDPR and requires \enquote{both at the time of the determination of the means for processing and at the time of the processing itself [(`at runtime')] [to] implement appropriate technical and organizational measures}. For the following considerations, we use the term \textit{inventory} to describe the aforementioned documentary measures concerning where personal data are being stored.

Through the technical lens, %
cloud native systems are built to scale applications for millions of concurrent users. For this, horizontal scaling techniques are used and infrastructure (compute, storage, and network) is provisioned automatically on demand \cite{gannon2017cloud}. Public cloud providers such as Amazon~Web~Services, Google~Cloud~Platform, Microsoft~Azure, or IBM~Cloud are appreciated for their elasticity and (seemingly) infinite resources. %
Consequently, privacy engineering has to examine this generation of technology intensively.

Moreover, the architectural paradigm of microservices is pervasive. Through cohesive, independently encapsulated services that interact through API-enabled messaging, scalable microservice architectures can be implemented \cite{dragoni2017microservices}. Within such microservices, the choice of data storage strongly depends on the functionality needed and the skills of the development team. Considered in its entirety, a multitude of polyglot microservices and multi-paradigmatic data stores are then in use. In larger distributed systems, we can observe thousands of such individual processing entities. From a transparency perspective, observing all of these concurrently for their (personal data) processing activities is a still-unsolved task. 

Besides these aspects, the degree of automation is rising. A best practice is to provide infrastructure as code (IaC) definitions. In particular, DevOps teams declaratively describe the desired state of compute, network, and storage components. Advanced configuration management %
and orchestration tools %
then transform the current state. For instance, an auto-scaling mechanism runs additional virtual machines or replicates a database under high load. From a privacy perspective, this continuous process -- being triggered by load, external events, and new code developments -- needs to be examined carefully. If a database instance was replicated to a new data center, possibly personal data were suddenly being stored in a third country. The GDPR would then require the controller to inventory this storage separately, implement adequate safeguards, and transparently inform data subjects.

\subsection{Data Loss Prevention} \label{ssec:dlp}
As mentioned above, Art. 5(1f) GDPR explicitly demands technical and organizational measures against accidental loss of personal data. %
This is why dedicated Data Loss Prevention systems (DLP) have been designed in several iterations. Most of these have in common that they examine \textit{data at rest}. Remarkably, \textit{data in transit} or \textit{data in use} will not be the core subject of this work.%
Existing DLP systems are mostly security-centric. %
Related work covers also user behavior analysis, such as profiling of document, database and network access \cite{shabtai2012data}, file and network traffic analysis \cite{li2015leakage} and protection from misuse of email communication \cite{alneyadi2016survey}, as well as content tagging for export prevention \cite{marecki2010decision} or access policies that hinder an adversary from accessing data stores. With DLP, the discovery of sensitive data has been demonstrated by the utilization of different NLP techniques, through which documents become classifiable \cite{trieu2017document}. Additionally, document classification has been discussed in the context of machine learning \cite{ghouse2019data}.%
Meanwhile, cloud providers developed proprietary DLP systems, namely there are AWS~Macie \cite{macie} and Google Cloud Data Loss Prevention \cite{google}. %
However, they are substantially limited to the extent they only support the provider's storage system (such as S3 for AWS) or they lack algorithmic transparency. Hence, multi-cloud and on-premise infrastructure are not covered at all. Neither meaningful evaluation establishing more trust has been published. Therefore, a data controller cannot meet the accountability requirements as imposed by, e.\,g., the GDPR \cite{kaul2021knowledge}.

In summary, the major drawback of these systems is the necessity to provide contextual information on where suspected data are located at. In microservice infrastructures, this step is far from easy. This is why we present a new general approach to identifying sensitive information, in particular personal data, with less necessary prior knowledge about the underlying system in the following sections.

\section{General Approach} \label{sec:general-approach}

\subsection{Requirements}

Hence, the system must be designed with a strong focus on the regulatory givens, which are in our example the information obligations from the GDPR.  To clarify up front what is often misunderstood: the discovery and inventory system is not required to store \textit{copies} of the personal data records persistently. Since the system is to be built to safeguard data protection rather than introducing novel threats, neither have the entities to be stored permanently in the inventory nor for analysis activities. Rather, the data minimization principle from Art. 5(1c) GDPR has to be ensured. This implies the system will only need to store meta data, e.g., database and records references. Thus, the system is to be built with a Privacy by Architecture \cite{spiekermann2008engineering} approach, not least to avert possible linkability attacks which arose from the presence of a personal record within the system \cite{pfitzmann2010terminology}.

Moreover, the system should cover a wide range of technological concepts, such as multi-paradigmatic infrastructures, storage alternatives, data types, etc., which all likely depend on the present infrastructure provider in a real cloud-native system. To support these concepts, the system shall be flexible and extensible with little effort.
The system must therefore interact through well-documented application programming interfaces (APIs) that power-efficient communication and promote the extensibility and connection with existing parts of the system or their development and operations (DevOps) tech stack.

Furthermore, multi-faceted automation potentials can unfold. These will help to replace existent laborious manual tasks and human errors (read: sending emails around, waiting for replies, and then manually creating spreadsheets that are outdated at the moment of their completion). The inventory process must therefore happen continuously and should scale out to larger infrastructures to meet the givens of current system architecture practice.

On a non-functional level, the system itself should be created and behave transparently to enable independent assessments, identify architectural and functional limitations, and determine
scalability,
accuracy,
security,
and usability -- also for a non-technical audience, since naturally multidisciplinary stakeholders are involved. %

\subsection{Introducing a Workflow Pattern for Scalable Discovery of Personal Data} \label{sec:workflow} %

When studying data loss prevention systems, there is a lot of attention on the detection methods but less on the practical integration of such systems. All too often, monolithic tools are proposed that are suitable to operate as standalone entity. However, these cannot meet the givens of modern heavily distributed systems consisting of numerous services, all potentially dealing with sensitive information. In these scenarios, we need to delegate the complexity of scheduling classification and inventorying tasks to dedicated algorithms. Therefore, we propose the application of the workflow pattern. Having also the requirements listed above in mind, we propose the following general four-step workflow as depicted in figure~\ref{fig:approach}:

\begin{figure}[h]
\includegraphics[width=1.0\linewidth]{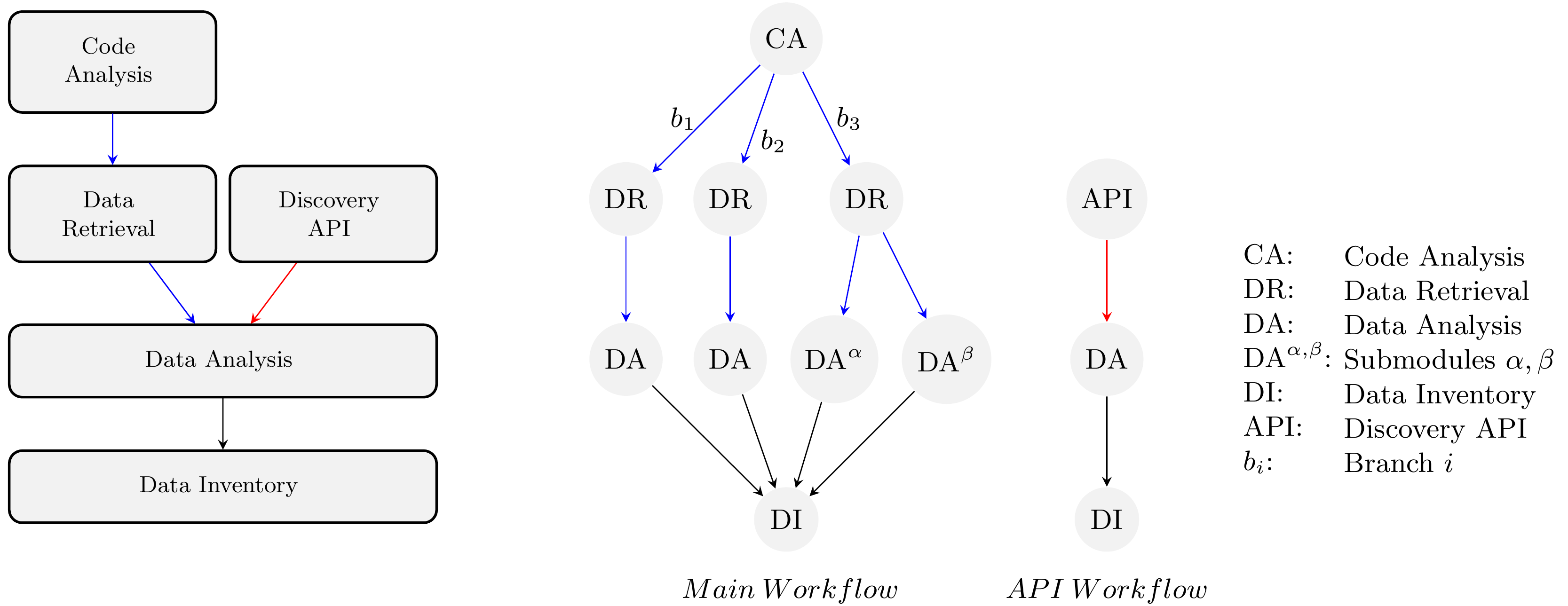}
\caption{Workflow pattern and exemplary workflows.}\label{fig:approach}
\end{figure}

\textbf{1. Code Analysis (CA): Discover storages in cloud native systems by analyzing version-controlled code repositories.} With this module, an approach to finding storages in unknown distributed systems environments shall be implemented (black-box approach). By analyzing code repositories, it is possible to find traces, definitions, or encoded connection strings of storages such as databases or disk storages. Repositories may contain the program's code, but also relevant Infrastructure as Code (IaC) definitions. Both of them potentially comprise information about personal data storages. The aspired outcome of the code analysis is a list of connection details for each storage found, which then enables the following retrieval component to read from. The list of connections must consist of a Uniform Resource Identifier (URI), login credentials if required, and the storage type. To successfully achieve the task of connection information extraction, the module would need to be able to clone code repositories and parse relevant code, including parameters of software packages used for storage connection (e.g., frequently used object-relational mappers). Incomplete connection information -- for instance in case of missing passwords or dynamically assigned IP  addresses or host names -- should be adequately marked in the inventory. Given these pieces of information, a data controller can take action to include them manually in future audits/analyses.

\textbf{2. Data Retrieval (DR): Manage to access the deployed storages and retrieve relevant (meta) data.} The Data Retrieval directly depends on the Code Analysis. With the given storage information, this component should be able to connect to the storages and fetch data and meta data from them. Structured and potentially unstructured data have to be extracted in the form of database entities or files. The whole process must stay reliable, and therefore a data subset selection (sampling) could optionally guarantee the timely termination of the following processing steps for very large data sets.
In addition, to provide a system that is applicable for as many scenarios and architectures as possible, a Discovery API needs to be added. This API listens for data requests, which will then be analyzed, to overcome the issue that storages were not found, or their connection information was incomplete. That includes external storages %
that are not under development of the data controller.
The implementation of middlewares which actively send data to the Discovery API allows the analysis of these prior hidden data collections. Alternatively, it may be used to analyze data that is not stored at rest, such as samples from stream processing frameworks.%

\textbf{3. Data analysis (DA): Detect personal data using heuristics and data analysis methods.} The Data Analysis component accepts data and meta data as incoming parameters. %
The actual analysis is carried out in a multistep approach, which results in indicators for a personal or non-personal data classification which includes entity and storage references to be inventoried. %
It is preferable to yield false-positive rather than false-negative classifications. For example, if verbose meta data -- which can be categorically similar to a personal field's data type, but also natural language such as field names -- exist, an analysis of the meta data can result in false-positive indicators. Primarily, sensitive information shall not be overlooked. It is safe to assume that these results are, before any optimizations, still preferred over overlooking sensitive information. Moreover, all analysis techniques which are convenient for meta data, especially pattern matching and lookup tables, should be also applied to data in the initial phase.

\textbf{4. Data Inventory (DI): Inventory findings while meeting the regulatory givens.} The results of all prior steps are then written into the Data Inventory which can serve as input for Data Protection Impact Assessments, Records of Processing Activities, or any other external or internal supervisory activity. Since the discovery job results can be highly diverse, a flexible and non-relational document store is considered the best fit here.
The results can be transformed to meet the form specifications of necessary legal documents, potential visualizations, or summaries of transparency information (Art. 12--14 GDPR).

Continuous software engineering relies on the concept of CI/CD pipelines, which are usually triggered by the version control system \cite{shahin2017continuous}. Therefore, most likely, the process shall be started each time relevant code changes are detected or due to changes in the underlying architecture of a system.

\subsection{Workflow engineering}

For large architectures and consequently many appearing triggering events, this four-step workflow needs to be properly orchestrated. In the probable event of detecting multiple storages under examination, the execution of the aforementioned steps needs then to be parallelized. Therefore, an efficient execution yielding timely results needs horizontal scalability. The components presented above are, in turn, dependent on each other. On a conceptual level, these dependencies can be modeled as a Directed Acyclic Graph (DAG). We therefore propose an orchestration implementing the \textit{workflow pattern} \cite{workflow}.
That means, each component performs a dedicated task for separation of concerns in an independent or dependent (and consequently sequential) flow.

Figure~\ref{fig:approach} depicts an example of an instantiated \textit{Main Workflow} with parallelization of the sub-graphs. Supposing the Code Analysis returns a list of three storages, the workflow is forked afterward into three execution branches. Then, the Data Retrieval component is instantiated in each of the branches and takes responsibility for the sub-tasks. For more complex scenarios, forking can even be nested to introduce even more parallelism. %
Assuming the Data Retrieval of branch $b_3$ has returned complex data that should be analyzed by multiple techniques, DA$^\alpha$ and DA$^\beta$ would fork the sub-graph to handle the data sets separately. Finally, all branches are joined and the workflow terminates after storing the findings successfully. Besides, the \textit{API Workflow} happens on requesting the Discovery API and could also be forked (not shown).%

In a cloud native environment, such workflows can be implemented through workflow management platforms \cite{workflow}. Their implementations depend on DAGs as they manage workflow orchestration (i.e., the concurrent and distributed task execution while meeting the ordering dependencies). Letting such a tool take the responsibility for task execution and scheduling of the discovery and inventory process promises reliability, scalability, and high automation potential, meeting the above-mentioned requirements. Keeping these aspects in mind, we now continue with the concrete software architecture and implementation in the upcoming sections.

\section{Software Architecture and Implementation} \label{sec:implementation}
We will now synthesize the conceptual workflow pattern and engineering considerations to elaborate on the design and implementation of a prototype system called \textsc{Teiresias}. The complete implementation is available under the MIT License as open source software in a public code repository.\footnote{
\url{https://github.com/teiresias-personal-data-discovery/teiresias-system}
}  

\subsection{Overview}

\usetikzlibrary{positioning, shadows, shapes.geometric, shapes.symbols, patterns}
\newcommand*\pointer[1]{\tikz[anchor=2mm]{\node[shape=circle,fill=black, text=white,scale=0.5] (char) {#1};}}

\begin{figure*}[!ht]
  \captionsetup{justification=centering, type=figure}
   \centering
   \includegraphics[width=1.0\linewidth]{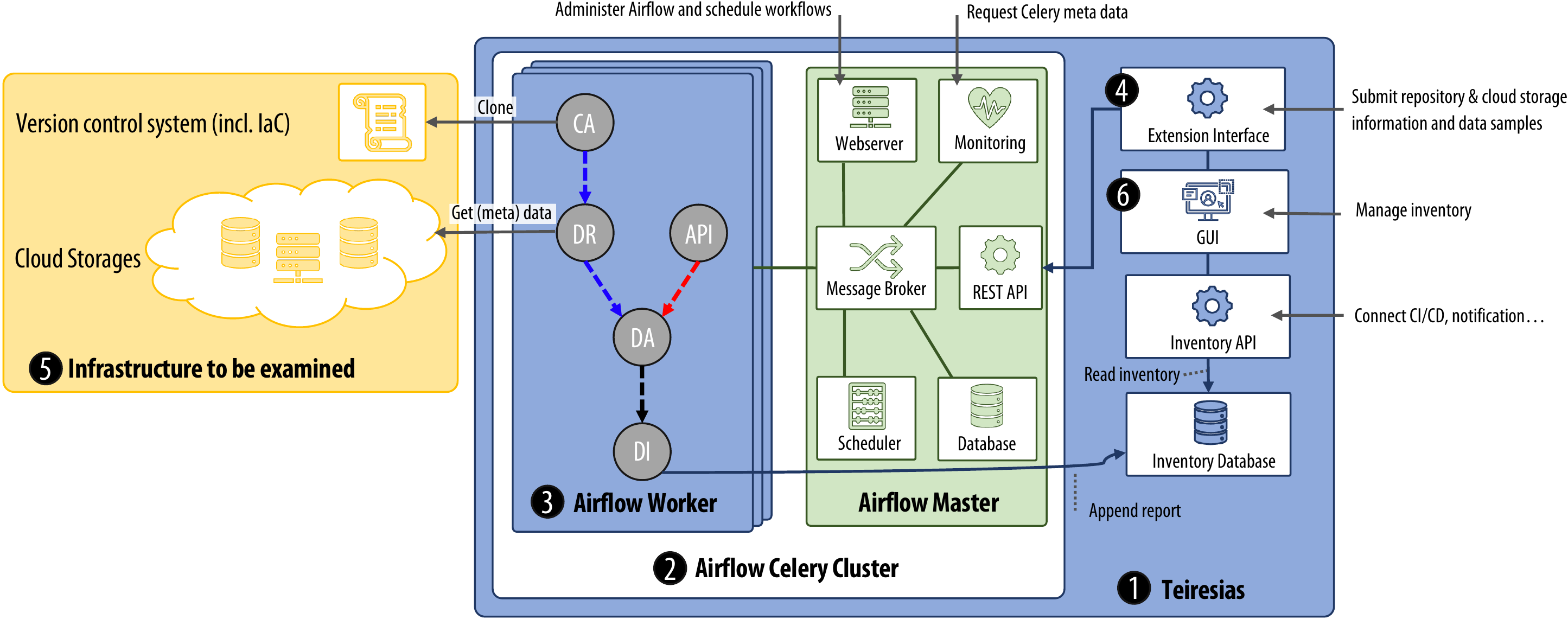}
   \caption{\label{fig:arch} System Architecture}
\end{figure*}

In figure~\ref{fig:arch} we provide an overview of the proposed system architecture. As indicated above, we heavily rely on the concepts of workflow management platforms. For our implementation we chose the open-source workflow platform Apache Airflow %
(for more details, see sec.~\ref{sec:airflow}). It is mainly responsible for the correct execution of all workflow steps. Within the \textsc{Teiresias} system boundary~\pointer{1}, Apache Airflow is integrated as an Airflow Celery Cluster~\pointer{2}, consisting of a message broker, database, (Airflow) API, scheduler, and worker nodes. An Airflow Worker~\pointer{3} is executing the code of each discovery component. Hence, it gets allocated their needed processing resources automatically. A data controller or supervisory authority can initialize the discovery and inventory process by passing (meta) data to the Extension Interface~\pointer{4}. Such data will then be passed on into the workflow, where modules are listening for incoming data (\emph{Discovery API}). Alternatively, they access them during a workflow execution, which may include fetching data from the data controller's infrastructure~\pointer{5} under examination through both the \emph{Code Analysis} and \emph{Data Retrieval} modules. Finally, discovered meta data are stored in the Inventory Database.
These findings can afterward be accessed through the Inventory GUI~\pointer{6}, which requests the reports from the Inventory Database through the Inventory API. Meanwhile, all the Airflow subcomponents can be administered through the Web Server (incl. GUI). For instance, workflows and rules for their scheduling, such as the determination of an examination on a regular basis, can be fine-tuned. Through these settings, \textsc{Teiresisas} can be precisely parameterized to fit the users' needs. 

\subsection{Workflow management} \label{sec:airflow}
We run the workflow management platform in a Celery cluster deployment. Celery handles tasks asynchronously based on a job queue, which communicates through message passing. We employ Redis as a broker to exchange messages between the scheduler and worker services. For all tasks, we implemented \emph{PythonOperators} for executing the logic of all components.%
The DAG for orchestrating the workflow is defined in Python. The definition of the DAG focused on scalability for the examination of larger infrastructures and extensibility for handling paradigmatically different services (e.g., different storages or IaC definitions). Conditional branching (as indicated in the example above) is introduced by placing a \emph{BranchPythonOperator} instance in front of the branches, which is chosen during DAG execution by conditional context evaluation. Moreover, parallelization of sub-workflows can be achieved by iterating over Airflow variables, and -- once per iteration -- the instantiation and linking (defining the execution order) of tasks are dynamically set. During runtime, the system would instantiate a \emph{clone\_and\_analyze\_code} function to examine a code repository for storage definitions and afterward create a task for each \emph{process\_code\_analysis} task, all of which can then run in parallel. %
Throughout the iterative prototypical implementation, stability could be improved by the re-implementation of the tasks by adding atomicity and idempotency as per best practice. For that, several more little DAG definitions (chunking them in smaller units) worked best. Failing tasks would then not prevent the main enclosing DAG to continue. Each error is logged for debugging and highlighted in the Airflow administration interface. We provide documentation along with the code.

\begin{figure*}[t]
  \captionsetup{justification=centering, type=figure}
   \centering
   \includegraphics[width=1.0\linewidth]{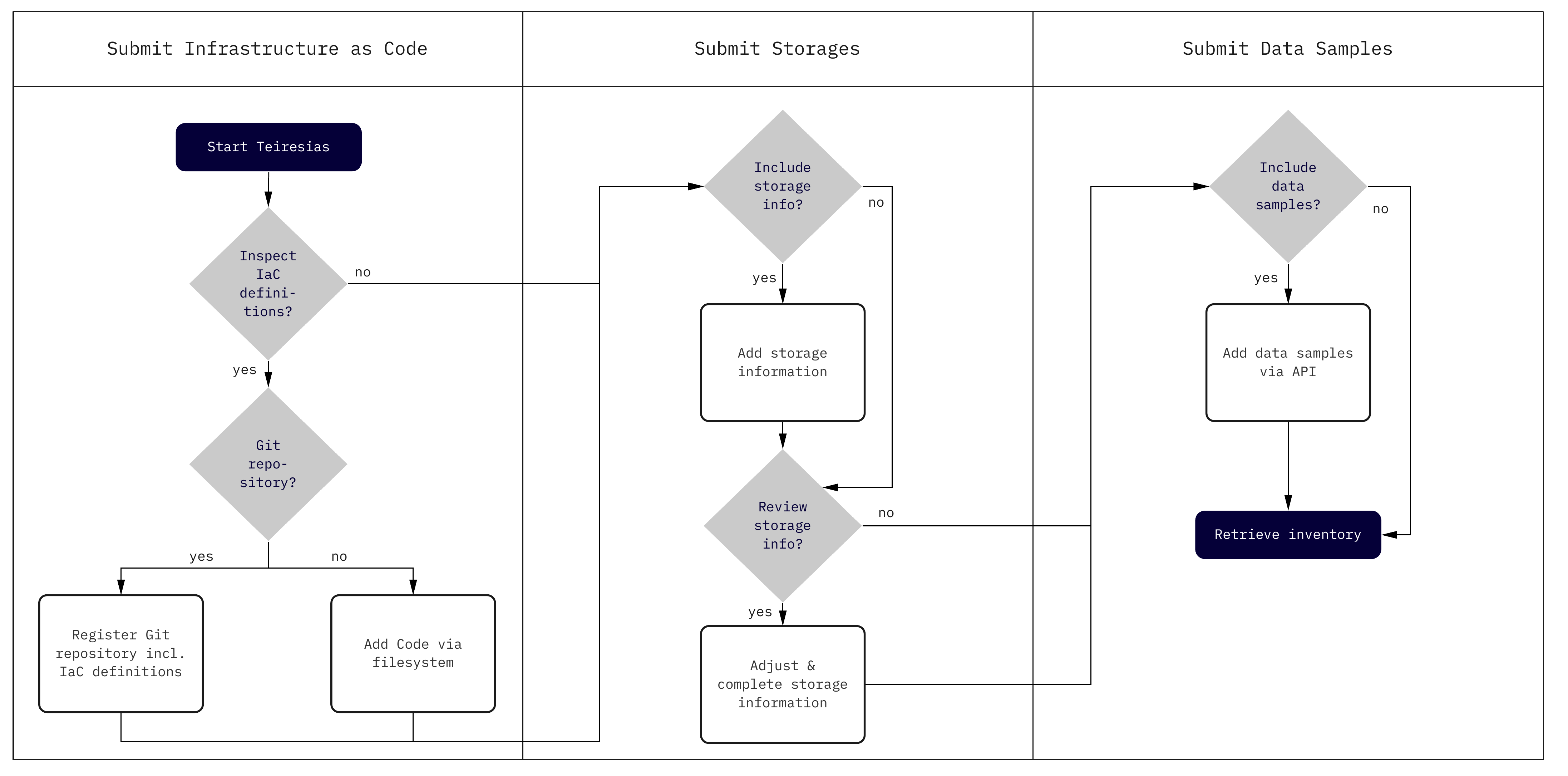}
   \caption{\label{fig:interaction} Interaction flow}
\end{figure*}

\subsection{Components}
We now briefly explain the implementation of the core components. We provide figure~\ref{fig:interaction} to show the basic interactions with the system.

\textbf{Code Analysis:} This component allows discovering and semantically parsing IaC definitions. In particular, we chose to support Ansible and Docker Compose encoded in YAML files. The analysis extracts, among others, details such as Docker images or Ansible module names. Through a look-up table and several storage \& IaC dependent regular expressions we identify, for demonstration purposes, PostgreSQL and MongoDB storages. There we fulfill the requirement of multi-paradigmatic storage identification. The analysis factors in structural information, retrieving the storages' contexts and resolves any variables that might be distributed over several IaC files. Moreover, the component can utilize repository information passed through the extension interface and clone Git repositories. Non-Git code bases can also be passed to the module by semi-automatically pulling the contents into a specific directory, in which a listener detects any changes and hence triggers the analysis. In doing so, (parts of) repositories can be analyzed without having access privileges to the version control system, which could be preferred from a security.%

\textbf{Data Retrieval and Analysis:} The component collects meta data about different data stores. At the core, an object-relational mapper (SQLAlchemy) allows for easy integration of a variety of storages, such as MySQL, Google BigQuery, or IBM DB2. While running, any analysis is performed \textit{in place} through the engines of the data storages. For example, the \textit{Meta data Analysis} of a PostgreSQL database fetches the list of columns per table, data types, number of entities, and primary key definitions. An \emph{INET}-typed table would then indicate the storage of IP addresses, which will often be considered relatable to a person. Throughout the analysis, a look-up table is compared to the attributes using simple named-entity recognition (NER) methods. For instance, the similarity is measured by the Levenshtein distance on a per attribute level. Finally, proximity lists are sorted and filtered using a 0.6 threshold, which is, according to our experiments, a viable trade-off between a low false-positive rate (i.e., excluding sub-string matches with few characters) and the best possible true positive rate (i.e., including matches of declined or compound words which hint at personal data semantically). Afterward, the \textit{Data Analysis} is implemented through regular expression-based search patterns (e.g., for social security, ID, or credit card numbers). The result is a list of references (e.g., \emph{ObjectId}s in a MongoDB) %
, which is appended to the report. Finally, the discovery comprises a binary classification of whether personal data have been detected. %
Since several meaningful insights originate from the analysis, such as the number of matched entities, the proximities from the meta data's analysis, and the total number of entities in the collection, should be weighted differently. For this purpose, we propose and implement the $T$ metric:

\begin{equation*}
\label{eq:word}
    \begin{split} 
  T&:=\underbrace{min(1, n)}_{\substack{=\text{ }0\text{ }\lor\text{ }1}}\cdot\underbrace{max(\overbrace{\alpha_{\text{hasMatch}}}^\text{= 0 $\lor$ 1}, \overbrace{\beta_{\text{meanProx}} / 100}^\text{= 0 $\lor$ x $\in$ [0.6, 1]})}_{\substack{\text{= 0 $\lor$ x $\in$ [0.6, 1]}}} \\
    n &:= \text{total number of entities in collection} \\
  \alpha_{\text{hasMatch}} &:= min(1, \text{total number of data matches})  \\
  \beta_{\text{meanProx}} &:= mean(\text{proximities of attribute names to keywords})  \\ \\
  \end{split}%
\end{equation*}

In short, we propose a binary classification based on $T$ which enables a quick assessment of data stores. However, we store the underlying measures because they are helpful for later possibly manual verification.
    
\textbf{Discovery API:} The API accepts any valid JSON document, which will then be analyzed. The pattern matching works the same way as described above. A report will be written with a user-provided identifier. Using the REST API of the Airflow web server, interoperability with other systems can be easily achieved.
    
\textbf{Data Inventory:} The reports are persistently stored in a MongoDB database, since the schemaless approach fits best to the potentially heterogeneous report structures. A specialized reporting Airflow operator is responsible for appending report portions from the different discovery steps into one bundled report per discovery execution. This is done by using a common DAG execution identifier, which is passed to the reporting operator. Each part is written, regardless of whether the following discovery steps are failing or succeeding, to provide full transparency and indicate necessary manual interaction.
    
\textbf{Extension Interface and GUI:} Furthermore, the inventory consists of a RESTful API and a graphical interface (GUI), through which the reports can be requested by the data controller or supervisory authority. Initially, the users register the repository URI and branch name, to start the workflow.
Next, additional storage information which was not accessible during execution can be completed. This includes the addition of externally managed data stores. Such information is then provided to the discovery via Airflow variables. Furthermore, storages can be temporarily excluded from the workflows or eventually be deleted. Such deletion is helpful when infrastructure has been redefined, and bypassing storages with huge amounts of known non-personal content can help to limit unnecessary processing costs. Such changes remain consistent between multiple code analysis executions.

\subsection{Deployment and Integration}

The deployment works as follows. First, all services are deployed through Docker containers. Referencing production-ready Docker images enable a quick, maintainable, repeatable, and deterministic deployment.%
Second, all critical system parts%
are only accessible after successful authentication. %
The communication with all accessible servers is encrypted using TLS, which was achieved by adding a reverse proxy server, which handles encryption with the utilization of a user-generated private key and certificate. The inventory database separates users for read and write access. It is additionally advised to provision a virtual private network to add another security layer. More extensive details are provided as software documentation in the repository.

Recommended integration scenarios include automated and recurrent triggering (e.g., hourly schedule) of the workflow through the management platform. Alternatively, the Discovery API can be triggered through CI/CD pipelines. Moreover, to align with DevPrivOps / DevSecOps \cite{devprivops, myrbakken2017devsecops} practices, the reports can be updated based on the workload. For example, each time the monitoring system registers changes, the inventory process should be triggered to check for unforeseen new sources of personal data.

\section{Evaluation} \label{sec:evaluation}

To evaluate our approach, we first compare our prototype to two widely used baseline systems, namely AWS Macie and Google CDLP, with regard to the performance of personal data discovery. Therefore, we compile three data sets to evaluate both (personal) data and meta data discovery in a single analysis task. Hence, we assume a single column of 5k values (i.e., potential personal data) with a meaningful column type and name (i.e., meta data). The first two data sets are composed of a synthetic personal data generator%
which outputs IP addresses using a regular expression and, through lookup tables, forenames, and surnames, which we combine (full names) to get a higher number of unique values. Besides these, we prepare a random subset of scraped personal handles from Twitter.\footnote{\url{https://github.com/danibram/mocker-data-generator},\\ \url{https://kaggle.com/hwassner/TwitterFriends}} %
Moreover, we create a set of labeled noise to not give the veneer of heuristics, since it would be feasible to drop empty data sets from the analysis queue in the systems. In particular, the labeled noise has four columns named \emph{user\_name, email, address} and \emph{ip}.
It is expected that \emph{IPv4} addresses can be discovered by pattern matching, whereas \emph{Full Names} are a use case for classification tasks or comprehensive lookup tables. Furthermore, \emph{Twitter Names} can only be found during classification and \emph{Labeled Noise} entities should not be detected as personal data at all.

Since AWS Macie exclusively scans files within its S3 storage, and Google CDLP does not cover non-proprietary databases,%
\emph{csv} files have been used for the cloud service evaluation%
Avoiding interference between the experiment iterations, each \emph{csv} file has been deployed individually to a freshly provisioned storage. For the prototype's evaluation, the same data sets have been deployed to different tables of a managed PostgreSQL database within the Google Cloud Platform (PostgreSQL~13, 2~vCPUs, 3.75~GB Memory).%
The Google CDLP parameter \emph{Percentage of included objects scanned within the bucket} was set to 100\% and the \emph{Sampling method} was set to \textit{No Sampling}. No other preferences have been set in AWS Macie's console since comparable defaults were set. In both the Google CDLP and the AWS Macie console, for each iteration, a one-time scan with a pointer to the specific bucket was submitted. The prototype ran on a MacBook Pro 2019 (2.4 GHz Intel i5 CPU, 16 GB RAM).

Table~1 summarizes the first experimental results. For both proprietary services, there is not an indication that meta data have been analyzed at all. Google CDLP has correctly found the 5\,000 \emph{IPv4} entities per regular expression. Our prototype is, in addition, able to find proximity between \emph{IPv4} and \emph{IP} through the meta data lookup attribute. There is some vagueness in the interpretation of which techniques have been used to classify the \emph{Full Name} entities correctly as personal data, which only AWS Macie was able to do. It is most likely that only pre-trained machine-learning models can achieve that task %
\textsc{Teiresias} again correctly classified the data set via the meta data analysis. \emph{Twitter Names} entities have not been recognized as personal data by any of the compared systems, but, the data set was classified by \textsc{Teiresias}' meta data analysis workflow. In turn, the cloud services have correctly classified the \emph{Labeled Noise} as non-personal data, and the \textsc{Teiresias} analysis classified it incorrectly as personal data, which is a result of the found meta data proximities and a non-empty data set. To overcome this false positive, the weighting of the terms of $T$ could be refined in future work. However, yielding false positives rather than false negatives rather strengthens the comprehensiveness of the discovery, since it prevents overlooking sensitive data. Balancing the results in a F1 measure, the proprietary services both reached a 0.57 score, each with one true positive, one true negative, and two false negatives. The prototype's score of 0.86 can be ascribed to the true positive classifications of the meta data analyses. %
In these first experiments, \textsc{Teiresias} outperforms AWS Macie and Google CDLP. Still, regarding the data analysis, only a section of personal data can be discovered with pattern matching and other rule-based detection mechanisms. Promising classification technologies %
should therefore be considered to be added to the system as future work.

In our second experiment, we measure the runtime performance, for which we re-used the experimental setup described above. However, this time we cannot compare AWS Macie and Google CDLP, since their underlying compute resources are not publicly known and this would contradict fairness in performance benchmarks.%
For the experiment, four different data sets with four columns each, and different total numbers of entities (0.5--500k) have been deployed. Afterward, we measured the execution time for the different samples. To limit the workload to one specific data set at a time, the PostgreSQL instance was registered to the system, containing one table per iteration, and the DAG for data analysis was scheduled exclusively.%

\begin{figure}
\includegraphics[width=\linewidth]{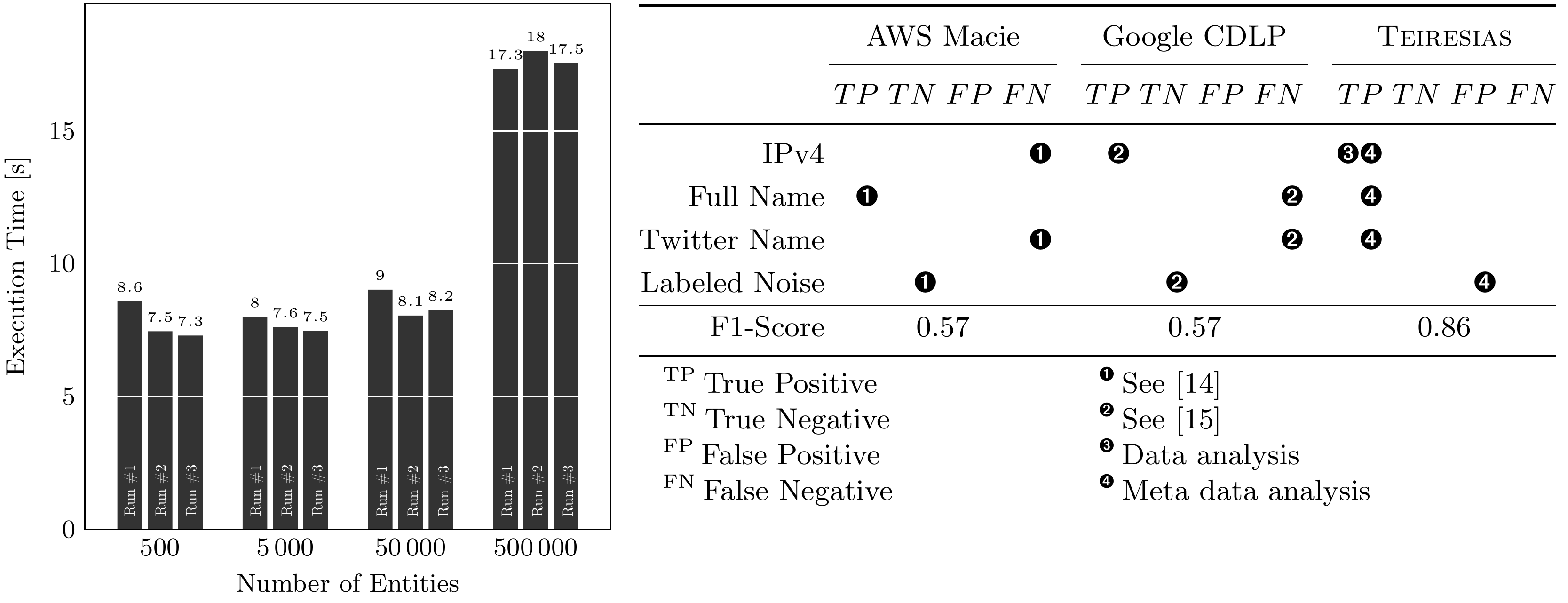}
\begin{floatrow}
\ffigbox{%
    }
{
      \caption{Execution times of three\\ runs for different sample sizes.}\label{fig:performance}
}
\capbtabbox[]{%
}{%
      \caption{Comparative evaluation results of DLP systems.}\label{table:evaluation}%
}
\end{floatrow}
\end{figure}

The results are presented in figure~3. We show that there is a non-linear growth, with 8.4 seconds for 50k and 17.6 seconds for 500k entities, which is nearly a doubling of the required time for a tenfold higher number of entities. It can be assumed that, with a very high number of entities, the mean execution time depends largely on the in-place data analysis query processing costs. In contrast, with a low number of entities, the costs for the system core functionalities and the meta data handling make up a major part of the execution time and stay almost constant. In summarizing, we demonstrate the applicability of our approach in a real-world setting, since these execution times allow for discovery operations without major time or resource consumption.

Note, we publish all test data sets used for the evaluation to enhance both repeatability and verifiability %
of our work.\footnote{%
\url{https://github.com/teiresias-personal-data-discovery/evaluation}
}  

\section{Limitations and Discussion} \label{sec:limitations}

Without a doubt, the system provides a scalable framework, which illustrates an automation concept using the proposed workflow pattern. However, for now, it only supports a limited set of data at rest storages and types of IaC definitions. To be used in more environments, support to connect to more popular \textit{at rest} systems should be implemented. More and more relevant processing happens also in stream processing systems and ephemeral storages, which could be included as well. Moreover, static code analysis techniques would complement the code analysis features to discover also some of these additional components. %
Furthermore, the detection and analysis methods should be extended. Lookup tables and regular expressions are well-functioning tools to detect many critical categories of personal data. Nevertheless, with the help of mature natural language processing and advanced machine learning models, better classification results can be yielded for all kinds of unpredictable data sets. This is especially important for audit scenarios and data that are provided by the data subject. This limitation, however, was not within the scope of the paper, but recent related work could complement our approach \cite{wei2022using}. %
Future work should also focus on data linkage attacks and anonymization or minimization methods \cite{pallas_icwe_2022_janus}. Since the GDPR requires data controllers to consider if it is reasonably likely (cf. Recital 26) that personal data get de-anonymized, a DLP system could also try to (re-)combine records from different storage systems. %

\textsc{Teiresias} is expected to discover and inventory personal data continuously at runtime. Related work shows how to formalize machine-readable transparency information (encoding the purpose, legal bases, third country transfers, categories of personal data, etc.)
\cite{grunewald2021tilt, gruenewald2021tira}. At the same time, observability measures (such as logging, distributed tracing, monitoring)
can create precise data flow models. Future research should concentrate on combining such practical privacy engineering practices to harvest accurate and transparent information at runtime \cite{sion2021overview, cloudinspector}. Especially growing and constantly evolving microservice architectures are not to be inventoried manually anymore. Therefore, there is an urgent need for more automated tools that help data controllers to keep track of their processing activities. %

\section{Conclusion} \label{sec:conclusion}

In this paper, we presented a DLP approach that monitors a system under examination for personal data at runtime continuously. %
With these contributions, data controllers are provided with working measures for aligning with regulatory frameworks such as the GDPR. Moreover, supervisory authorities can utilize \textsc{Teiresias} to audit infrastructures. %
Practical applicability has been demonstrated through a comparative evaluation. We emphasize that the approach primarily targets cloud native systems, but is also applicable to cloud-enabled ones. The latter, e.g., in legacy-cloud hybrids, would only need a lightweight custom middleware component to be connected to the proposed APIs. %
Within the wider prospects, our workflow could be extended to not only detect personal data (and other kinds of sensitive information) but also support efficient distributed data deletion.%

\subsubsection*{Acknowledgements.}
\begin{small} The work behind this paper was partially conducted within the project DaSKITA, supported under grant no. 28V2307A19 by funds of the Federal Ministry for the Environment, Nature Conservation, Nuclear Safety and Consumer Protection (BMUV) based on a decision of the Parliament of the Federal Republic of Germany via the Federal Office for Agriculture and Food (BLE) under the innovation support programme. \end{small}

\bibliographystyle{IEEEtran}
\bibliography{biblio}

\end{document}